\documentclass{PoS}

\usepackage[sort&compress,numbers]{natbib}
\usepackage{mathtools}
\usepackage{amsfonts} 
\usepackage{amssymb} 
\usepackage{amsmath} 
\usepackage{graphicx} 
\usepackage{latexsym} 
\usepackage{verbatim} 

\newcommand{\Kdf}[0]{\mathcal K_{\mathrm{df},3}}

\newcommand{\Mth}[0]{\mathcal M_{ 3}}
\newcommand{\ML}[0]{\mathcal M_{ 2,L}}
\newcommand{\MthL}[0]{\mathcal M_{ 3,L}}
\newcommand{\K}[0]{\mathcal K_2}
\newcommand{\M}[0]{\mathcal M_2}

\title{Progress on the three-particle quantization condition}

\ShortTitle{Progress on the three-particle quantization condition}

\author{Ra\'ul A. Brice\~no\\
Thomas Jefferson National Accelerator Facility,
12000 Jefferson Avenue, Newport News, VA 23606, USA\\
        E-mail: \email{rbriceno@jlab.org}}

\author{Maxwell T. Hansen\\ 
 Institut f\"ur Kernphysik and Helmholz Institute Mainz, Johannes Gutenberg-Universit\"at Mainz, 
55099 Mainz, Germany\\
        E-mail: \email{hansen@kph.uni-mainz.de}}

\author{\speaker{Stephen R. Sharpe}\\
        Physics Department, University of Washington, Seattle, WA 98195-1560, USA\\
        E-mail: \email{srsharpe@uw.edu}}

\abstract{We report progress on extending the relativistic model-independent quantization condition 
for three particles, derived previously by two of us, to a broader class of theories, 
as well as progress on checking the formalism. 
In particular, we discuss the extension to include the possibility of 2->3 and 3->2 transitions 
and the calculation of the finite-volume energy shift of an Efimov-like three-particle bound state.
The latter agrees with the results obtained previously using non-relativistic quantum mechanics (NRQM).}

\FullConference{The 34th International Symposium on Lattice Field Theory,\\
		24-30 July, 2016\\
		University of Southampton, UK}

\begin{document}

\section{Introduction}

Tremendous progress has been made over the last few years in the
calculation of resonance properties from first principles using lattice QCD (LQCD).
The present frontier is the determination of the
properties of resonances coupling to multiple two-body channels.
A recent example is the study of resonances in Ref.~\cite{Dudek:2016cru}.
This considers resonances coupling to both the $\pi\eta$ and $K\overline K$ channels,
albeit with quarks that are heavier than physical.
It uses a theoretical formalism---a ``two-particle quantization condition" 
generalized from seminal papers by L\"uscher~\cite{Luscher:1986pf,Luscher:1991cf}---that
relates the spectrum in a finite volume (FV) to the infinite-volume scattering amplitudes.

However, as the quark mass is lowered, an increasing number of resonances
couple either dominantly, or in part, to three-particle channels.
Examples include $\omega\to3\pi$, $K^*\to K\pi\pi$ and $N(1440)\to N\pi\pi$
as well as the recently discovered $X$, $Y$ and $Z$ resonances.
If we wish to determine the properties of such resonances from
first principles, it is essential to have a generalization of the
two-particle quantization condition to one that includes three particles.
Such an extension is also needed to use LQCD to study weak decays involving
three particles, e.g. $K\to 3\pi$.

In Refs.~\cite{SpectoK,KtoM} two of us have provided such a generalization,
applicable to three identical, relativistic, spinless particles whose interactions 
are constrained by a G-parity-like symmetry. We briefly describe this work,
referring to Refs.~\cite{SpectoK,KtoM} for details.
It consists of two parts. The first is a three-particle quantization condition\footnote{%
This result holds up to exponentially suppressed FV effects, proportional to
$e^{-mL}$, that we ignore throughout.}
\vspace{-0.25cm}
\begin{equation}
\det(F_3^{-1} + \Kdf) = 0\,.
\label{eq:QC3}
\end{equation}
where here and below all quantities are infinite-dimensional matrices
in the space of on-shell three-particle states in FV.\footnote{%
In any practical application this matrix space must be truncated, as in the two-particle case.}
$\Kdf$ is a three-particle generalization of the K-matrix---an infinite-volume
quantity that is, however, not physical as it contains an UV cutoff.
$F_3$ is the matrix
\begin{equation}
F_3 = \frac{F_2}{2\omega L^3}
\left[ \frac13 - \ML F_2 - \mathcal D^{(u,u)}_L \frac{F_2}{2\omega L^3}\right]\,,
\label{eq:F3def}
\end{equation}
where $L$ is the box size (assuming a cubic box), $\omega$ the relativistic energy,
$F_2$ is a generalized L\"uscher zeta-function (a known volume-dependent matrix),
$\ML$ is a FV version of the two-particle scattering amplitude,
and $\mathcal D^{(u,u)}_L$ is the contribution to the FV three-particle scattering
amplitude that contains only two-particle interactions.
The key point is that $F_3$ depends only on known, $L$-dependent
kinematic functions and the infinite-volume two-particle scattering amplitude $\M$.
Thus it can be determined by applying the two-particle quantization condition to
the two-particle FV spectrum. 

The second part of the three-particle formalism connects $\Kdf$ to
the infinite-volume three-particle scattering amplitude $\Mth$~\cite{KtoM}.
The latter is obtained using
\begin{equation}
\Mth = \lim_{L\to\infty}\big|_{i\epsilon} \MthL\,,
\ \ {\rm with} \ \ 
\MthL = \mathcal S \left[ \mathcal D^{(u,u)}_L
+ \mathcal L_L^{(u)} \Kdf \frac1{1+F_3 \Kdf} \mathcal R_L^{(u)}\right]\,.
\label{eq:M3res}
\end{equation}
Here the $i\epsilon$ subscript indicates a particular infinite-volume limit, and $\MthL$
is a FV version of the scattering amplitude. 
The quantities $\mathcal L_L^{(u)}$ and $\mathcal R_L^{(u)}$
depend on $\M$ and known kinematic functions, like $F_3$ and $\mathcal D^{(u,u)}_L$.
Thus if $\M$ and $\Kdf$ are obtained, respectively, from the two- and three-particle
quantization conditions, then $\Mth$ can, in principle, be determined from 
Eq.~(\ref{eq:M3res}). Working out the details, one finds that this
requires solving nested UV-finite integral equations involving on-shell quantities
[arising from the implicit matrix indices in Eq.~(\ref{eq:M3res})]~\cite{KtoM}.

There are two major limitations of this formalism. First, it assumes a $Z_2$ symmetry forbidding
$1\to2$, $2\to 3$, etc. transitions. Thus it is applicable (to good approximation) to three pions,
where G-parity enforces the $Z_2$ symmetry, but not to most three-particle systems.
Second, it requires that the two-particle channel be nonresonant in the kinematic range
of interest. For example, in a three-pion system with vanishing total momentum,
angular-momentum and $I=1$, the total energy must satisfy $E < m_\pi + m_R$, 
where $m_R$ is the position of the pole in the two-particle K-matrix corresponding
to the lightest $f_0$ resonance.
This is a very serious limitation on the practical applicability of the formalism.\footnote{%
Other restrictions---to identical, and thus necessarily degenerate, particles and to 
spinless particles---are expected to be simpler to remove,
based on experience with two particles. We do not discuss these here.}

\section{Extensions of the formalism}

We are actively working on removing the two limitations just described, and provide
a brief update on the status of this work. We have made the most progress on removing the
$Z_2$ symmetry. Based on our analysis so far,
we conjecture that the generalized quantization condition is
\begin{equation}
\det\left[
\left(\begin{array}{cc} F_2^{-1} & 0 \\ 0 & F_3^{-1} \end{array} \right)
+
\left(\begin{array}{cc} \K & \mathcal K_{3\to2} \\\mathcal K_{2\to3} & \Kdf \end{array} \right)
\right] = 0\,.
\label{eq:QC3noZ2}
\end{equation}
This rather natural generalization of the three-particle quantization condition (\ref{eq:QC3}),
and the corresponding two-particle result $\det(F_2+\K^{-1})=0$,
extends the matrix indices to contain both two- and three-particle on-shell FV phase space.
This extension arises from the fact that correlators in the $Z_2$-less theory have cuts
containing of any number of on-shell particles. The result (\ref{eq:QC3noZ2})
holds for $m < E < 4m$, where only two and three-particle cuts are allowed.
The key point is that physical, on-shell $2\to 3$ and $3\to 2$ transitions are allowed,
and this leads to the off-diagonal terms in the second matrix in (\ref{eq:QC3noZ2}).
These off-diagonal terms contain infinite-volume K-matrices that, like $\Kdf$ 
(and $\K$ below threshold),  are unphysical.

One key feature of this conjectured result is that, unlike $\Kdf$, the $2\to3$ and $3\to2$ K-matrices do not
contain divergences arising from long-distance propagation of on-shell particles. 
$\mathcal K_{2\to3}$  and $\mathcal K_{3\to2}$  are quasi-local vertices that can be
expanded in spherical harmonics.

To establish Eq.~(\ref{eq:QC3noZ2}) requires extending the analysis of Ref.~\cite{SpectoK}.
One begins with the skeleton expansion of a FV correlator, locates the position of all possible
power-law FV dependence (which are the two- and three-particle cuts), and then replaces
FV momentum sums with integrals plus the difference. The skeleton expansion here is
more complicated than with a $Z_2$ symmetry, 
requiring a large number of additional Bethe-Salpeter kernels. 
Nevertheless, we have a partial argument leading to (\ref{eq:QC3noZ2}), and hope
to complete it soon. 

Having done so, the second step will be to relate the four infinite-volume-but-unphysical
K-matrices $\{\K,\mathcal K_{2\to3},\mathcal K_{3\to2},\Kdf\}$ to
the infinite-volume scattering amplitudes $\{\M,\mathcal M_{2\to3},\mathcal M_{3\to2},\Mth\}$,
i.e. to generalize Eq.~(\ref{eq:M3res}).
Our preliminary results indicate that the resulting integral equations 
are coupled, so that all four K-matrices must be known at a given energy in order
to determine any of the scattering amplitudes.

We are at an earlier stage in removing
the second limitation of the original formalism, 
namely the requirement that $\K$ have no above-threshold poles.
Our approach is to use the factorization of the residues of these poles to simplify
the resulting expressions, and to explicitly account for the new FV effects that these
poles introduce.
\vspace{-.25cm}

\section{A new test of the formalism: FV energy shift for a three-particle bound state }

In the remainder of this talk I report on a new test of the original three-particle
formalism of Refs.~\cite{SpectoK,KtoM}. This is based on work with Hansen 
that is now written up in Ref.~\cite{HSMRR}. I present here only an overview of the argument.

We think that it is important to provide checks of the formalism, since it is rather involved
and required a very lengthy derivation. We have already completed one such check
in Refs.~\cite{thresh,HSpert}, where we have determined the energy of the three-particle
state closest to threshold in a series in $1/L$. The $1/L^3$, $1/L^4$ and $1/L^5$ terms
agree, as expected, with results from NRQM.
The $1/L^6$ term contains relativistic effects, and is also the first term in which the three-particle
scattering amplitude enters.
We have done an auxiliary calculation of the threshold energy 
in relativistic $\lambda\phi^4$ theory, 
working through ${\cal O }(\lambda^3)$ in perturbation theory~\cite{HSpert}.
The results for the $1/L^{3-6}$ terms from our formalism and perturbation theory 
are in complete agreement.

The new check presented here is based on the work of Ref.~\cite{MRR},
hereafter referred to as MRR. These authors
use NRQM to determine the leading volume dependence of the energy of a
three-particle bound state with total momentum $\vec P=0$. 
Specifically, they assume only two-particle
potentials, and that these are near the unitary limit of infinite scattering length.
In this limit, first considered by Efimov~\cite{Efimov}, there is a sequence of
three-particle bound states.\footnote{%
This is true for both signs of the scattering length. Here we assume a positive
scattering length so that there are no two-body bound states.}
Focusing on one such bound state, MRR find
\begin{align}
E_B &= 3m -\kappa^2/m + \Delta E_3(L) \,,
\\
\Delta E_3(L) &= c \frac{\kappa^2}{m} \frac1{(\kappa L)^{3/2}}
\exp\left( - 2\kappa L/\sqrt3\right)
 \left[1 +{\cal O}\left(\frac{\kappa}{m},\frac{1}{\kappa L} \right) \right]
\,.
\label{eq:MRR}
\end{align}
The first equation defines $\kappa$, while the second gives the leading
volume dependence of the energy. The constant $c$ is known, and depends on
the detailed form of the wavefunction of the (infinte-volume) Efimov state.\footnote{%
Here we use a slightly different definition of $c$ than that used by MRR or in Ref.~\cite{HSMRR}.}
Our aim here is to fully reproduce Eq.~(\ref{eq:MRR}) using our formalism.

The corresponding equation for a two-particle bound state can be determined
from L\"uscher's original quantization condition (and from NRQM),
and takes the form~\cite{Beane:2003da}
\begin{equation}
\Delta E_2(L) = -12 \frac{\kappa^2}{m} \frac1{\kappa L} e^{-\kappa L}
 \left[1 +{\cal O}\left(\frac{\kappa}{m},\frac{1}{\kappa L} \right) \right]
 \,.
 \label{eq:DE2}
 \end{equation}
 Thus we see that the three-particle case has a different exponent, different power of $1/L$,
 and a more complicated constant.
 
The calculation of MRR assumes that only s-wave interactions contribute.
Making this approximation in our formalism
reduces the size of the matrix space in Eqs.~(\ref{eq:QC3}) and (\ref{eq:M3res}).
The matrix index is now given solely by the momentum of one of the three particles---the
``spectator"---while the other two are in a relative s-wave.\footnote{%
In FV this spectator momentum is quantized. Note that
the asymmetry inherent in the choice of spectator is removed 
by subsequent symmetrization in our formalism.}
In all subsequent formulae the spectator momentum is shown explicity,
and there are no implicit matrix indices.

The logic of the calculation is straightforward.\footnote{%
The following description is an updated version of that given in the talk, based on
subsequent improvements in the derivation. In particular, we no longer
need to make the approximation $\Kdf=0$.}
$\Mth$ has, by assumption, a pole at $E = E_B$:
\begin{equation}
\Mth(\vec p, \vec k) = -
\frac{\Gamma(\vec p)\overline\Gamma(\vec k)}{E^2-E_B^2}
+ \textrm{non-pole}
\,,
\label{eq:pole}
\end{equation}
where the residues are the amputated, on-shell Bethe-Salpeter amplitudes.
We also know that, since $\MthL$  is a FV correlation function,
it has poles at the energies of the FV states, and in particular at the shifted FV energy of
the bound state.
Given Eq.~(\ref{eq:M3res}), we know that, for large $L$, $\MthL$ is close to $\Mth$.
The idea is then to quantify this closeness 
by systematically determining the volume dependence of 
$\MthL$ using Eq.~(\ref{eq:M3res}).

$\Mth$ and $\MthL$ differ because momentum integrals in the former are replaced by
sums (i.e. by matrix products) in the latter. We can systematically replace sums with
integrals plus sum-integral differences. 
Since we are working below threshold, with no on-shell intermediate states,
the volume dependence of the sum-integral differences is exponentially suppressed, 
but with the exponent proportional to $\kappa L$. 
This leads to a much weaker suppression than that from
the contributions proportional to $e^{-mL}$  that we consistently neglect.
We keep the terms with the smallest exponential suppression, which turn out to have
exponent $-2\kappa L/\sqrt3$, and also drop terms suppressed by additional powers of $1/L$.
Using the properties of the residue functions $\Gamma$ and $\overline\Gamma$, 
which will be discussed below, 
we find that the dominant FV corrections arise from three-particle cuts lying between
the scattering of one pair and a different pair (which we label ``switch states").
With this simplification we are able to show that
the unsymmetrized versions of $\MthL$ and $\Mth$ satisfy\footnote{%
Unsymmetrized means that the first two-particle interaction occurring in a skeleton expansion of
$\Mth$ and $\MthL$ occurs between the non-spectator pair. $\M(\vec \ell)$ gives the
scattering amplitude for two particles with total four-momentum 
$P_2=(E-\omega_\ell,-\vec \ell)$. In the sum, $\vec\ell= 2\pi \vec n/L$ with $\vec n$ 
a vector of integers.}
\begin{equation}
\MthL^{(u,u)}(\vec p, \vec k) = \Mth^{(u,u)}(\vec p, \vec k)
+ \left[\frac1{L^3}\sum_{\vec \ell} - \int \frac{d^3\ell}{(2\pi)^3} \right]
\Mth^{(u,u)}(\vec p,\vec \ell) \frac1{2\omega_\ell \M(\ell)}
\MthL^{(u,u)}(\vec \ell, \vec k) \,.
\label{eq:MthLfromMth}
\end{equation}
The unsymmetrized amplitudes have poles at the same positions as the symmetrized ones,
and by substituting the pole forms [e.g. Eq.~(\ref{eq:pole})] into this equation we find the
energy shift to be
\begin{equation}
\Delta E_3(L) = - \frac1{2E_B} \left[\frac1{L^3}\sum_{\vec k} - \int \frac{d^3k}{(2\pi)^3}\right]
\frac{\overline \Gamma^{(u)}(k) \Gamma^{(u)}(k)}{2\omega_k \M(k)}
\,.
\label{eq:DEHS}
\end{equation}
Here $\Gamma^{(u)}$, $\overline\Gamma^{(u)}$ are the residues
appearing in the pole form for the unsymmetrized amplitude $\Mth^{(u,u)}$.
Note that these turn out to depend only on the magnitude of the spectator momentum,
as does $\M$.

To proceed, we need to know the residue functions, which, as noted above, are
the amputated, on-shell versions of the Bethe-Salpeter (BS) amplitudes for the bound state.
We know the Schr\"odinger wavefunction of the bound states (reviewed, for example,
in Ref.~\cite{BH}), so the issue is how to obtain the BS amplitudes from the wavefunction.
This question was addressed, long ago, in Ref.~\cite{FF}.
This work assumed only instantaneous two-particle interactions and worked in the 
nonrelativistic limit. These are the assumptions made also by MRR, so the result
of Ref.~\cite{FF} is sufficient for our purposes.
Using this result (which we have checked in detail, as a full derivation is not supplied in
Ref.~\cite{FF}, and corrected the normalization factor) we find\footnote{%
The same result holds for $\overline\Gamma^{(u)}$ aside from complex conjugation.
This schematic form does not show the arguments of $\Gamma^{(u)}$ and $\widetilde\phi_3$,
as we do not have space to explain all the details here. For these see Ref.~\cite{HSMRR}.}
\begin{equation}
\Gamma^{(u)} =  4\sqrt 3 m^2  \lim_{\rm on\ shell} 
\left[-\frac{\kappa^2}{m} - \sum_{i=1,3} \frac{\vec p_i^2}{2m}\right] \widetilde \phi_3
\,.
\end{equation}
$\tilde\phi_3$ is the Fourier transform of the part of the wavefunction
that corresponds to the unsymmetrized BS amplitude. It satisfies the Fadeev equation,
in which only the potential between two of the particles appears.
The key point here is that the explicit form of $\phi_3$ is known~\cite{BH}.
The meaning of ``on shell" is that the free Schr\"odinger operator,
i.e. the term in square brackets, vanishes. The right-hand side does not vanish, however,
because $\widetilde\phi_3$ diverges. We find
\begin{equation}
\overline\Gamma^{(u)}(k)\Gamma^{(u)}(k) =
-c \ 64\ 3^{3/4}\pi^{5/2} \frac{m^2}{\kappa^2} 
\left[1 + \frac{3k^2}{4\kappa^2}\right]^{-1} + \dots
\,,
\label{eq:Gammares}
\end{equation}
with $c$ the same constant as in Eq.~(\ref{eq:MRR}).
Here we have kept only the leading singularity for small $k$, since this leads to the
dominant FV correction when inserted into Eq.~(\ref{eq:DEHS}).

To evaluate the Eq.~(\ref{eq:DEHS}), we also need $\M$ in the unitary limit:
\begin{equation}
\frac1{\M(k)} = \frac{\kappa}{32\pi m} \left[1 + \frac{3k^2}{4\kappa^2}\right]^{1/2} 
\,.
\end{equation}
Inserting this and Eq.~(\ref{eq:Gammares}) into (\ref{eq:DEHS}), using the
Poisson summation formula, and evaluating the integral, we find
\begin{equation}
\Delta E_3(L) = c \frac{\kappa^2}{m} \frac{2}{3^{1/4}\sqrt\pi \kappa L} 
K_1\left(\frac{2\kappa L}{\sqrt3}\right)\,,
\end{equation}
whose asymptotic form agrees with the MRR result, Eq.~(\ref{eq:MRR}).

\section{Extension to a moving three-particle bound state}

Within our formalism, it is straightforward to generalize this result to a moving frame.
In the two-particle case, the corresponding generalization of the rest-frame result,
Eq.~(\ref{eq:DE2}),
has been given in Ref.~\cite{Davoudi:2011md}. 
They found, for total momentum $\vec P=2\pi \vec n_P/L$, the simple form 
\begin{equation}
\Delta E_{2,\vec P}(\vec L) = f_2[\vec n_P] \Delta E_2(L)\,,\qquad
f_2[\vec n_P] = \tfrac16 \sum_{\hat s} e^{i 2\pi \hat s\cdot \vec n_P/2}\,,
\end{equation}
where the sum runs over the six integer vectors of unit length. 
Thus the form of the volume dependence is unchanged, but the overall factor depends
on $\vec P$. This dependence turns out to be rather dramatic,
e.g. $f_2[\vec 0]=1$ while $f_2[(1,1,1)]=-1$.
Furthermore, by combining results from different frames,
the leading exponential dependence can be canceled~\cite{Davoudi:2011md}.

We find a similar form in the three-particle case, but with a different prefactor:
\begin{equation}
\Delta E_{3,\vec P}(\vec L) = f_3[\vec n_P] \Delta E_3(L)\,,\qquad
f_3[\vec n_P] = \tfrac16 \sum_{\hat s} e^{i 2\pi \hat s\cdot \vec n_P/3}\,.
\end{equation}
Again the prefactor various rapidly with momentum. The fact that the volume dependence
changes only by an overall factor means that, as in the two-particle case, it is possible
to cancel the leading dependence by combining results from different frames.
In fact, since $f_3[(1,1,0)]=0$,
the volume dependence for $\vec n_P=(1,1,0)$ is subleading.

\section{Acknowledgments}
We thank Akaki Rusetsky for helpful discussions.
RAB acknowledges support from U.S. Department of Energy contract DE-AC05-06OR23177, 
under which Jefferson Science Associates, LLC, manages and operates Jefferson Lab.
SRS was supported in part by the United States Department of Energy 
grant DE-SC0011637.


\bibliographystyle{apsrev4-1}
\bibliography{ref}

\begin{thebibliography}{14}%
\makeatletter
\providecommand \@ifxundefined [1]{%
 \@ifx{#1\undefined}
}%
\providecommand \@ifnum [1]{%
 \ifnum #1\expandafter \@firstoftwo
 \else \expandafter \@secondoftwo
 \fi
}%
\providecommand \@ifx [1]{%
 \ifx #1\expandafter \@firstoftwo
 \else \expandafter \@secondoftwo
 \fi
}%
\providecommand \natexlab [1]{#1}%
\providecommand \enquote  [1]{``#1''}%
\providecommand \bibnamefont  [1]{#1}%
\providecommand \bibfnamefont [1]{#1}%
\providecommand \citenamefont [1]{#1}%
\providecommand \href@noop [0]{\@secondoftwo}%
\providecommand \href [0]{\begingroup \@sanitize@url \@href}%
\providecommand \@href[1]{\@@startlink{#1}\@@href}%
\providecommand \@@href[1]{\endgroup#1\@@endlink}%
\providecommand \@sanitize@url [0]{\catcode `\\12\catcode `\$12\catcode
  `\&12\catcode `\#12\catcode `\^12\catcode `\_12\catcode `\%12\relax}%
\providecommand \@@startlink[1]{}%
\providecommand \@@endlink[0]{}%
\providecommand \url  [0]{\begingroup\@sanitize@url \@url }%
\providecommand \@url [1]{\endgroup\@href {#1}{\urlprefix }}%
\providecommand \urlprefix  [0]{URL }%
\providecommand \Eprint [0]{\href }%
\providecommand \doibase [0]{http://dx.doi.org/}%
\providecommand \selectlanguage [0]{\@gobble}%
\providecommand \bibinfo  [0]{\@secondoftwo}%
\providecommand \bibfield  [0]{\@secondoftwo}%
\providecommand \translation [1]{[#1]}%
\providecommand \BibitemOpen [0]{}%
\providecommand \bibitemStop [0]{}%
\providecommand \bibitemNoStop [0]{.\EOS\space}%
\providecommand \EOS [0]{\spacefactor3000\relax}%
\providecommand \BibitemShut  [1]{\csname bibitem#1\endcsname}%
\let\auto@bib@innerbib\@empty
\bibitem [{\citenamefont {Dudek}\ \emph {et~al.}(2016)\citenamefont {Dudek},
  \citenamefont {Edwards},\ and\ \citenamefont {Wilson}}]{Dudek:2016cru}%
  \BibitemOpen
  \bibfield  {author} {\bibinfo {author} {\bibfnamefont {J.~J.}\ \bibnamefont
  {Dudek}}, \bibinfo {author} {\bibfnamefont {R.~G.}\ \bibnamefont {Edwards}},
  \ and\ \bibinfo {author} {\bibfnamefont {D.~J.}\ \bibnamefont {Wilson}}
  (\bibinfo {collaboration} {Hadron Spectrum}),\ }\href {\doibase
  10.1103/PhysRevD.93.094506} {\bibfield  {journal} {\bibinfo  {journal} {Phys.
  Rev.}\ }\textbf {\bibinfo {volume} {D93}},\ \bibinfo {pages} {094506}
  (\bibinfo {year} {2016})},\ \Eprint {http://arxiv.org/abs/1602.05122}
  {arXiv:1602.05122 [hep-ph]} \BibitemShut {NoStop}%
\bibitem [{\citenamefont {Luscher}(1986)}]{Luscher:1986pf}%
  \BibitemOpen
  \bibfield  {author} {\bibinfo {author} {\bibfnamefont {M.}~\bibnamefont
  {L\"uscher}},\ }\href {\doibase 10.1007/BF01211097} {\bibfield  {journal}
  {\bibinfo  {journal} {Commun. Math. Phys.}\ }\textbf {\bibinfo {volume}
  {105}},\ \bibinfo {pages} {153} (\bibinfo {year} {1986})}\BibitemShut
  {NoStop}%
\bibitem [{\citenamefont {Luscher}(1991)}]{Luscher:1991cf}%
  \BibitemOpen
  \bibfield  {author} {\bibinfo {author} {\bibfnamefont {M.}~\bibnamefont
  {L\"uscher}},\ }\href {\doibase 10.1016/0550-3213(91)90584-K} {\bibfield
  {journal} {\bibinfo  {journal} {Nucl. Phys.}\ }\textbf {\bibinfo {volume}
  {B364}},\ \bibinfo {pages} {237} (\bibinfo {year} {1991})}\BibitemShut
  {NoStop}%
\bibitem [{\citenamefont {Hansen}\ and\ \citenamefont
  {Sharpe}(2014)}]{SpectoK}%
  \BibitemOpen
  \bibfield  {author} {\bibinfo {author} {\bibfnamefont {M.}~\bibnamefont
  {Hansen}}\ and\ \bibinfo {author} {\bibfnamefont {S.}~\bibnamefont
  {Sharpe}},\ }\href {\doibase 10.1103/PhysRevD.90.116003} {\bibfield
  {journal} {\bibinfo  {journal} {Phys.Rev.}\ }\textbf {\bibinfo {volume}
  {D90}},\ \bibinfo {pages} {116003} (\bibinfo {year} {2014})},\ \Eprint
  {http://arxiv.org/abs/1408.5933} {arXiv:1408.5933 [hep-lat]} \BibitemShut
  {NoStop}%
\bibitem [{\citenamefont {Hansen}\ and\ \citenamefont {Sharpe}(2015)}]{KtoM}%
  \BibitemOpen
  \bibfield  {author} {\bibinfo {author} {\bibfnamefont {M.}~\bibnamefont
  {Hansen}}\ and\ \bibinfo {author} {\bibfnamefont {S.}~\bibnamefont
  {Sharpe}},\ }\href {\doibase 10.1103/PhysRevD.92.114509} {\bibfield
  {journal} {\bibinfo  {journal} {Phys. Rev.}\ }\textbf {\bibinfo {volume}
  {D92}},\ \bibinfo {pages} {114509} (\bibinfo {year} {2015})},\ \Eprint
  {http://arxiv.org/abs/1504.04248} {arXiv:1504.04248 [hep-lat]} \BibitemShut
  {NoStop}%
\bibitem [{\citenamefont {Hansen}\ and\ \citenamefont
  {Sharpe}(2016{\natexlab{a}})}]{HSMRR}%
  \BibitemOpen
  \bibfield  {author} {\bibinfo {author} {\bibfnamefont {M.~T.}\ \bibnamefont
  {Hansen}}\ and\ \bibinfo {author} {\bibfnamefont {S.~R.}\ \bibnamefont
  {Sharpe}},\ }\href@noop {} {\  (\bibinfo {year} {2016}{\natexlab{a}})},\
  \Eprint {http://arxiv.org/abs/1609.04317} {arXiv:1609.04317 [hep-lat]}
  \BibitemShut {NoStop}%
\bibitem [{\citenamefont {Hansen}\ and\ \citenamefont
  {Sharpe}(2016{\natexlab{b}})}]{thresh}%
  \BibitemOpen
  \bibfield  {author} {\bibinfo {author} {\bibfnamefont {M.}~\bibnamefont
  {Hansen}}\ and\ \bibinfo {author} {\bibfnamefont {S.}~\bibnamefont
  {Sharpe}},\ }\href {\doibase 10.1103/PhysRevD.93.096006} {\bibfield
  {journal} {\bibinfo  {journal} {Phys. Rev.}\ }\textbf {\bibinfo {volume}
  {D93}},\ \bibinfo {pages} {096006} (\bibinfo {year} {2016}{\natexlab{b}})},\
  \Eprint {http://arxiv.org/abs/1602.00324} {arXiv:1602.00324 [hep-lat]}
  \BibitemShut {NoStop}%
\bibitem [{\citenamefont {Hansen}\ and\ \citenamefont
  {Sharpe}(2016{\natexlab{c}})}]{HSpert}%
  \BibitemOpen
  \bibfield  {author} {\bibinfo {author} {\bibfnamefont {M.}~\bibnamefont
  {Hansen}}\ and\ \bibinfo {author} {\bibfnamefont {S.}~\bibnamefont
  {Sharpe}},\ }\href {\doibase 10.1103/PhysRevD.93.014506} {\bibfield
  {journal} {\bibinfo  {journal} {Phys. Rev.}\ }\textbf {\bibinfo {volume}
  {D93}},\ \bibinfo {pages} {014506} (\bibinfo {year} {2016}{\natexlab{c}})},\
  \Eprint {http://arxiv.org/abs/1509.07929} {arXiv:1509.07929 [hep-lat]}
  \BibitemShut {NoStop}%
\bibitem [{\citenamefont {Meissner}\ \emph {et~al.}(2015)\citenamefont
  {Meissner}, \citenamefont {Rios},\ and\ \citenamefont {Rusetsky}}]{MRR}%
  \BibitemOpen
  \bibfield  {author} {\bibinfo {author} {\bibfnamefont {U.-G.}\ \bibnamefont
  {Mei{\ss}ner}}, \bibinfo {author} {\bibfnamefont {G.}~\bibnamefont {R\'ios}}, \
  and\ \bibinfo {author} {\bibfnamefont {A.}~\bibnamefont {Rusetsky}},\ }\href
  {\doibase 10.1103/PhysRevLett.117.069902, 10.1103/PhysRevLett.114.091602}
  {\bibfield  {journal} {\bibinfo  {journal} {Phys. Rev. Lett.}\ }\textbf
  {\bibinfo {volume} {114}},\ \bibinfo {pages} {091602} (\bibinfo {year}
  {2015})},\ \bibinfo {note} {[Erratum: Phys. Rev.
  Lett.117,no.6,069902(2016)]},\ \Eprint {http://arxiv.org/abs/1412.4969}
  {arXiv:1412.4969 [hep-lat]} \BibitemShut {NoStop}%
\bibitem [{\citenamefont {Efimov}(1970)}]{Efimov}%
  \BibitemOpen
  \bibfield  {author} {\bibinfo {author} {\bibfnamefont {V.}~\bibnamefont
  {Efimov}},\ }\href {\doibase 10.1016/0370-2693(70)90349-7} {\bibfield
  {journal} {\bibinfo  {journal} {Phys. Lett.}\ }\textbf {\bibinfo {volume}
  {B33}},\ \bibinfo {pages} {563} (\bibinfo {year} {1970})}\BibitemShut
  {NoStop}%
\bibitem [{\citenamefont {Beane}\ \emph {et~al.}(2004)\citenamefont {Beane},
  \citenamefont {Bedaque}, \citenamefont {Parreno},\ and\ \citenamefont
  {Savage}}]{Beane:2003da}%
  \BibitemOpen
  \bibfield  {author} {\bibinfo {author} {\bibfnamefont {S.~R.}\ \bibnamefont
  {Beane}}, \bibinfo {author} {\bibfnamefont {P.~F.}\ \bibnamefont {Bedaque}},
  \bibinfo {author} {\bibfnamefont {A.}~\bibnamefont {Parre\~no}}, \ and\
  \bibinfo {author} {\bibfnamefont {M.~J.}\ \bibnamefont {Savage}},\ }\href
  {\doibase 10.1016/j.physletb.2004.02.007} {\bibfield  {journal} {\bibinfo
  {journal} {Phys. Lett.}\ }\textbf {\bibinfo {volume} {B585}},\ \bibinfo
  {pages} {106} (\bibinfo {year} {2004})},\ \Eprint
  {http://arxiv.org/abs/hep-lat/0312004} {arXiv:hep-lat/0312004 [hep-lat]}
  \BibitemShut {NoStop}%
\bibitem [{\citenamefont {Braaten}\ and\ \citenamefont {Hammer}(2006)}]{BH}%
  \BibitemOpen
  \bibfield  {author} {\bibinfo {author} {\bibfnamefont {E.}~\bibnamefont
  {Braaten}}\ and\ \bibinfo {author} {\bibfnamefont {H.-W.}\ \bibnamefont
  {Hammer}},\ }\href {\doibase 10.1016/j.physrep.2006.03.001} {\bibfield
  {journal} {\bibinfo  {journal} {Phys. Rept.}\ }\textbf {\bibinfo {volume}
  {428}},\ \bibinfo {pages} {259} (\bibinfo {year} {2006})},\ \Eprint
  {http://arxiv.org/abs/cond-mat/0410417} {arXiv:cond-mat/0410417} \BibitemShut
  {NoStop}%
\bibitem [{\citenamefont {Feldman}\ and\ \citenamefont {Fulton}(1982)}]{FF}%
  \BibitemOpen
  \bibfield  {author} {\bibinfo {author} {\bibfnamefont {G.}~\bibnamefont
  {Feldman}}\ and\ \bibinfo {author} {\bibfnamefont {T.}~\bibnamefont
  {Fulton}},\ }\href {\doibase 10.1016/0550-3213(82)90048-7} {\bibfield
  {journal} {\bibinfo  {journal} {Nucl. Phys.}\ }\textbf {\bibinfo {volume}
  {B195}},\ \bibinfo {pages} {61} (\bibinfo {year} {1982})}\BibitemShut
  {NoStop}%
\bibitem [{\citenamefont {Davoudi}\ and\ \citenamefont
  {Savage}(2011)}]{Davoudi:2011md}%
  \BibitemOpen
  \bibfield  {author} {\bibinfo {author} {\bibfnamefont {Z.}~\bibnamefont
  {Davoudi}}\ and\ \bibinfo {author} {\bibfnamefont {M.~J.}\ \bibnamefont
  {Savage}},\ }\href {\doibase 10.1103/PhysRevD.84.114502} {\bibfield
  {journal} {\bibinfo  {journal} {Phys.Rev.}\ }\textbf {\bibinfo {volume}
  {D84}},\ \bibinfo {pages} {114502} (\bibinfo {year} {2011})},\ \Eprint
  {http://arxiv.org/abs/1108.5371} {arXiv:1108.5371 [hep-lat]} \BibitemShut
  {NoStop}%
\end{thebibliography}%

\end{document}